\documentclass{ws-procs9x6}

\begin{document}

\title{Aspects of Neutrino Mass Matrices}

\author{P.H. FRAMPTON}

\address{Department of Physics and Astronomy, \\
University of North Carolina at Chapel Hill, \\ 
Chapel Hill, NC 27599-3255.\\ 
E-mail: frampton@physics.unc.edu}


\maketitle

\abstracts{After an Introduction briefly describing
the rise and fall of the three-zero texture of the Zee model,
we discuss still-allowed two-zero textures for the Majorana three-neutrino
mass matrix. Finally, a model with two right-handed neutrinos
and two Dirac texture zeros is described (FGY model)
which can relate CP violation in leptogenesis to CP
violation in long-baseline neutrino oscillations. }

\section{Introduction}

The minimal standard model involves three chiral neutrino states, but
it does not admit renormalizable interactions that
can generate neutrino masses. Nevertheless, experimental
evidence show that
both solar and atmospheric neutrinos display flavor oscillations,
and  hence that neutrinos do have mass.

Among the many renormalizable and gauge-invariant extensions of the standard model
that can provide neutrino masses are:

\begin{itemize}

\item The introduction of a complex triplet of mesons
($T^{++},\, T^+,\,T^0$) coupled bilinearly to pairs of lepton
doublets\cite{LSWMSW}.
They must also couple bilinearly to the
Higgs doublet(s) so as to avoid spontaneous $B-L$ violation and the
appearance of a massless and experimentally excluded  majoron.
 This mechanism can generate an arbitrary  complex symmetric
Majorana mass matrix for neutrinos.

\item The introduction of singlet counterparts to the
neutrinos with very large Majorana masses. The interplay between these mass
terms and those generated by the Higgs boson---the so-called
see-saw mechanism---yields an arbitrary
but naturally small Majorana neutrino mass matrix.

\item The introduction of a charged singlet meson $f^+$
coupled antisymmetrically to pairs of lepton doublets, {\it and\/} a
doubly-charged singlet meson $g^{++}$ coupled bilinearly both to pairs of
lepton singlets and to pairs of $f$-mesons. An arbitrary Majorana neutrino
mass matrix is generated in two loops.

\item The introduction  of a charged singlet meson $f^+$
coupled antisymmetrically to pairs of lepton doublets {\it and\/} (also
antisymmetrically) to a pair of Higgs doublets.  This
simple mechanism was first proposed by Zee~\cite{Zee},
and
results (at one loop) in a Majorana mass matrix
in the flavor basis ($e,\,\mu,\,\tau$)
 of a  special form:

\begin{equation}
{\bf M}= \left( \begin{array}{ccc}
0&m_{e\mu}&m_{e\tau} \\
m_{e\mu}&0&m_{\mu\tau} \\
m_{e\tau}&m_{\mu\tau}&0
\end{array}
\right)
\end{equation}
Because the diagonal entries of {\bf M} are zero,  the amplitude for
no-neutrino double beta decay vanishes at lowest order
and this process cannot proceed at an observable rate.
Furthermore, the
parameters $m_{e\mu},\, m_{e\tau}$ and
$m_{\mu\tau}$ may be taken real and non-negative
without loss of generality, whence  {\bf M} becomes  real as well as
traceless and
symmetric.  With this convention, the analog to the
Kobayashi-Maskawa matrix becomes orthogonal:
{\bf M} is explicitly
CP invariant and $\delta=0$. 

\end{itemize}

The sum of the neutrino masses (the eigenvalues of {\bf M})
vanishes: $ m_1+m_2+m_3=0 $.
When the squared-mass hierarchy
is taken into account, that the solar squared mass difference 
is very much smaller than the atmospheric squared mass difference,
there are two possibilities.
In case A, we have
$m_1+m_2 = 0$ and $m_3= 0$. This case arises
iff at least one of
the three parameters in {\bf M} vanishes.
In case B, we have $m_1= m_2$  and  $m_3= -2m_2>0$. 
It has been shown\cite{FG,FOY} that
neither case is consistent with the SuperKamiokande
and SNO data so the the Zee model is ruled out
by these experiments.

\bigskip
\bigskip

\section{Two-Zero Textures}

Having excluded one three-zero texture we now study
two-zero possibilities.

Quark masses and mixings are described by two $3\times
3$ matrices which involve a total of ten convention-independent real
quantities: four Kobayashi-Maskawa parameters and six positive quark
masses. The lepton sector requires only three additional parameters (the
charged lepton masses) in the minimal---and manifestly incomplete---standard
model wherein neutrinos are massless.

We assume that neutrino phenomenology can be formulated in
terms of three left-handed neutrino states with a complex symmetric Majorana
mass matrix ${\bf M}$. We do not commit ourselves to any particular mechanism
by which neutrino masses are generated. Of the twelve real parameters
characterizing ${\bf M}$, three are arbitrary phases of the three flavor
eigenstates. Thus the neutrino sector involves nine convention-independent
parameters. The two squared-mass differences and the four Kobayashi-Maskawa
analogs are likely to be measureable in practice. So also is the quantity
$\vert {\bf M}_{ee}\vert$ through the search for neutrinoless double beta
decay. There remain two parameters that are measureable in principle but
apparently not easily in practice. 

It will be useful to define the ratio of squared-mass differences, whose
estimated value is:
$R_\nu \equiv {\Delta_s\over \Delta_a}\approx  2\times 10^{-2}$.

We turn to the question of which
two independent entries of ${\bf M}$ can vanish in the basis wherein the
charged lepton mass matrix is diagonal.
 Of the fifteen logical
possibilities, we find   just seven to be
 in accord with our empirical hypotheses.  We discuss them
individually, with the
non-vanishing
entries  in each case denoted by $X$'s. Our results are presented to leading
order in the small parameter $s_2$.
 We begin with a texture in which ${\bf M}_{ee}={\bf M}_{e\mu}=0$:

\noindent{Case $A_1$: ~~
$ \pmatrix{0&0&X\cr 0&X&X\cr X&X&X\cr}$}
\medskip

\noindent
For this texture,
 ${\bf M}_{ee}=0$ so that  the amplitude for no-neutrino double beta
 decay vanishes to lowest order in neutrino masses. If Case $A_1$
 is realized in nature,
 the neutrinoless process simply cannot be detected. There is even more to
 say. Two of the squared neutrino masses
are suppressed relative to the third by a factor of $s_2^2$.
As a result
 we find that
 $s_2$ can lie  close  to
  its present experimental upper limit. This prediction will become more
precise when $\theta_3$ is better measured.
However,
 the CP-violating parameter $\delta$ is entirely
unconstrained. For case $A_1$  the subdominant
angle $\theta_2$ is likely to be
 measureable  and neutrinos may display observable  CP violation.
\bigskip\medskip

\noindent{Case $A_2$: ~~
$ \pmatrix{0&X&0\cr X&X&X\cr 0&X&X\cr}$}

\medskip

\noindent  
This texture has ${\bf M}_{ee}={\bf M}_{e\tau}=0$.  
Its phenomenological consequences are nearly the same  as those of Case $A_1$.

\bigskip\medskip

\noindent{Case $B_1$: ~~
$ \pmatrix{X&X&0\cr X&0&X\cr 0&X&X\cr}$}

\medskip

\noindent 
With  ${\bf M}_{\mu\mu}={\bf M}_{e\tau}=0$, we find
 an acceptable solution if and only if
$\vert s_2 \,\cos{\delta}\tan{2\theta_1}\vert \ll 1$, in which case
the three neutrinos are nearly degenerate
 in magnitude because $t_1^2\simeq 1$.
The texture is also promising in regard to
the search
for  neutrinoless double beta decay. We obtain
${\bf M}_{ee} \simeq  -t_1^2 \sqrt{\Delta_a/|1-t_1^4|}$
likely to exceed 100~meV.
The rate of neutrinoless double beta  decay may  approach
its current experimental upper limit.

\medskip
\bigskip
\medskip

\noindent{Case $B_2$: ~~
$ \pmatrix{X&0&X\cr 0&X&X\cr X&X&0\cr}$}

\medskip

\noindent  

This texture, with  ${\bf M}_{\tau\tau}={\bf M}_{e\mu}=0$, is
as in $B_1$ with $t_1$ replaced by $-1/t_1$. Its
phenomenological consequences are nearly the same  as those of Case $B_1$.
\bigskip\medskip

\noindent{Case $B_3$:  
$ \pmatrix{X&0&X\cr 0&0&X\cr X&X&X\cr}$ \qquad  and \qquad Case $B_4$: ~~
$ \pmatrix{X&X&0\cr X&X&X\cr 0&X&0\cr}$}
\medskip

\noindent
The phenomenological implications of all four cases $B_i$
(i = 1, 2, 3, 4) are substantially
the same.

\noindent {Case $C$: ~~
$ \pmatrix{X&X&X\cr X&0&X\cr X&X&0\cr}$}

\medskip

\noindent 
Our seventh and last  allowed texture has
 ${\bf M}_{\mu\mu}={\bf M}_{\tau\tau}=0$.

No other two-zero texture of the neutrino mass matrix is compatible with our
empirical hypotheses. It is easily verified that no two of our allowed
two-zero textures can  be simultaneously
satisfied while remaining  consistent with our
empirical hypotheses.
It follows  that there is no tolerable
three-zero texture.

The seven  allowed two-zero neutrino textures  fall into three classes:
$A$ (with two members), $B$ (with four members), and $C$.
 The textures within  each class are  difficult
or impossible to distinguish
experimentally, but each of the three classes has
radically different implications.
For class $A$, the subdominant angle
$\theta_2$ is expected to be relatively large, but  no-neutrino
$\beta\beta$ decay  is forbidden. For class $B$,  the latter process
should be measureable by the next generation of double beta decay
experiments, whilst $s_2$ may or may not be large enough to be detected.
However, if $s_2$ is comparable to its experimental upper limit, CP violation
must be nearly maximal and should be readily detectable by
proposed experiments.  For
class  $C$,  no-neutrino
$\beta\beta$ decay is likely to be observable  and
$\theta_2$ ought to  be large enough to
measure and to permit a search for CP violation.
It is surprising that such a great variety of textures of the neutrino mass
matrix can fit what is presently known about neutrino masses and
oscillations.
Future data should reveal
which, if any, of these textures should serve as a guide to the model
builder.

\medskip

\section{FGY Model}

One of the most profound ideas is\cite{Sakharov}
that baryon number asymmetry arises in the early universe
because of processes which violate CP symmetry and that
terrestrial experiments on CP violation
could therefore inform us of the details of such
cosmological baryogenesis.

The early discussions of baryogenesis focused on
the violation of baryon number and its possible relation to
proton decay. In the light of present evidence for neutrino masses
and oscillations
it is more fruitful to associate the baryon number
of the universe with violation of lepton number\cite{FY}.
In the present Letter
we shall show how, in one class of models, the sign of the
baryon number of the universe correlates with
the results of CP violation in neutrino oscillation
experiments which will be performed in the forseeable
future.

Present data on atmospheric and solar neutrinos suggest
that there are respective squared mass differences
$\Delta_a \simeq 3 \times 10^{-3} eV^2$
and
$\Delta_s \simeq 5 \times 10^{-5} eV^2$.
The corresponding mixing angles $\theta_1$
and $\theta_3$ satisfy
$tan^2 \theta_1 \simeq 1$
and $0.6 \leq sin^2 2\theta_3 \leq 0.96$
with $sin^2 \theta_3 = 0.8$ as the best fit.
The third mixing angle is much smaller than the other
two, since the data require $sin^2 2 \theta_2 \leq 0.1$.

A first requirement is that our model\cite{FGY} accommodate
these experimental facts at low energy.

In the minimal standard model, neutrinos are massless.
The most economical addition to the standard model
which accommodates both neutrino masses
and allows the violation
of lepton number to underly the cosmological baryon asymmetry
is two right-handed neutrinos $N_{1,2}$.

These lead to new terms in the lagrangian:

\begin{eqnarray}
{\bf L} & = & \frac{1}{2} (N_1, N_2) \left(
\begin{array}{cc} M_1 & 0 \\ 0 & M_2
\end{array} \right)
\left( \begin{array}{c} N_1 \\ N_2 \end{array} \right)
+  \nonumber \\
& + & (N_1, N_2)
\left( \begin{array}{ccc} a  &  a^{'}  &  0  \\
0  &  b  &  b^{'}  \end{array} \right)
\left( \begin{array}{c} l_1  \\  l_2 \\ l_3 \end{array}
\right)  H  + h.c.
\label{Lag}
\end{eqnarray}
where we shall denote the rectangular Dirac mass matrix
by $D_{ij}$. We have assumed a texture
for $D_{ij}$ in which the upper
right and lower left entries vanish.
The remaining parameters in our model
are both necessary and sufficient
to account for the data.

For the light neutrinos, the see-saw mechanism leads to
the mass matrix\cite{Y}
\begin{eqnarray}
\hat{L} & = & D^T M^{-1} D \nonumber \\
& = & \left( \begin{array}{ccc}
\frac{a^2}{M_1}  &  \frac{a a^{'}}{M_1}  &  0  \\
\frac{a a^{'}}{M_1}  &  \frac{(a^{'})^2}{M_1} + \frac{b^2}{M_2}
&  \frac{b b^{'}}{M_2} \\
0  &  \frac{b b^{'}}{M_2}  &  \frac{(b^{'})^2}{M_2}
\end{array}  \right)
\label{L}
\end{eqnarray}

We take a basis where $a, b, b^{'}$ are real and where
$a^{'}$ is complex $a^{'} \equiv |a^{'}|e^{i \delta}$.
To check consistency with low-energy
phenomenology we temporarily take the specific
values (these will be loosened later)
$b^{'} = b$ and $a^{'} = \sqrt{2} a$ and all parameters real.
In that case:
\begin{eqnarray}
\hat{L}
& = & \left( \begin{array}{ccc}
\frac{a^2}{M_1}  &  \frac{\sqrt{2}a^2}{M_1}  &  0  \\
\frac{\sqrt{2}a^2}{M_1}  &  \frac{2a^2}{M_1} + \frac{b^2}{M_2}
&  \frac{b^{2}}{M_2} \\
0  &  \frac{b^{2}}{M_2}  &  \frac{b^{^2}}{M_2}
\end{array}  \right)
\label{LL}
\end{eqnarray}
We now diagonalize to the mass basis by writing:
\begin{equation}
{\bf L} = \frac{1}{2} \nu^T \hat{L} \nu
= \frac{1}{2} \nu^{'T} U^T \hat{L} U \nu^{'}
\end{equation}
where
\begin{eqnarray}
U & = & \left( \begin{array}{ccc} 1/\sqrt{2}  &  1/\sqrt{2}  &  0  \\
- 1/2 & 1/2  &  1/\sqrt{2}  \\
1/2  &  -1/2 &  1/\sqrt{2}
\end{array} \right) \times \nonumber \\
& \times &
\left( \begin{array}{ccc}
1  &  0  &  0  \\
0  &  cos\theta &  sin\theta  \\
0  &  - sin\theta  &  cos\theta
\end{array}
\right)
\end{eqnarray}
We deduce that the mass eigenvalues  and $\theta$
are given by

\begin{equation}
m(\nu_3^{'}) \simeq 2 b^2/M_2; ~~~ m(\nu_2^{'}) \simeq 2 a^2 /M_1; ~~~
m(\nu_1^{'}) = 0
\end{equation}
and
\begin{equation}
\theta \simeq m(\nu_2^{'}) / (\sqrt{2} m(\nu_3^{'}))
\end{equation}
in which it was assumed that $a^2/M_1 \ll b^2/M_2$.

By examining the relation between the three mass eigenstates
and the corresponding flavor eigenstates
we find
that for the unitary matrix relevant to neutrino oscillations
that
\begin{equation}
U_{e3} \simeq sin\theta/\sqrt{2} \simeq m(\nu_2)/(2m(\nu_3))
\end{equation}

Thus the assumptions $a^{'} = \sqrt{2} a$, $
b^{'} = b$ adequately fit the
experimental data, but
$a^{'}$ and $b^{'}$ could
be varied around
$\sqrt{2}a$ and $b$ respectively
to achieve better fits.

But we may conclude that
\begin{eqnarray}
2b^2/M_2 & \simeq & 0.05 eV = \sqrt{\Delta_a} \nonumber \\
2a^2/M_1 & \simeq & 7 \times 10^{-3} eV  = \sqrt{\Delta_{s}}
\label{numass}
\end{eqnarray}

It follows from these values that $N_1$ decay satisfies the
out-of-equilibrium condition for leptogenesis (the
absolute requirement is $m < 10^{-2} eV$ \cite{buchpascos})
while $N_2$ decay does not.
This fact enables us to predict
the sign of CP violation in neutrino oscillations without ambiguity.

Let us now come to the main result. Having
a model consistent with all low-energy data and with
adequate texture zeros\cite{FGM} in $\hat{L}$ and
equivalently $D$ we can
compute the sign both of the high-energy
CP violating parameter ($\xi_H$) appearing
in leptogenesis and of the CP violation parameter
which will be measured in low-energy
$\nu$ oscillations ($\xi_L$).

We find the baryon number $B$ of the universe
produced by $N_1$ decay
proportional to\cite{Buch}
\begin{eqnarray}
B & \propto & \xi_H =
(Im D D^{\dagger} )_{12}^2 = Im (a^{'} b)^2 \nonumber \\
& = & + Y^2a^2b^2 sin 2\delta
\label{highenergy}
\end{eqnarray}
in which $B$ is positive by observation of the universe.
Here we have loosened our assumption about $a^{'}$
to $a^{'} = Y a e^{i \delta}$.

At low energy the CP violation in neutrino oscillations is governed by
the quantity
\begin{equation}
\xi_L = Im (h_{12} h_{23} h_{31})
\end{equation}
where $h = \hat{L} \hat{L}^{\dagger}$.

Using Eq.(\ref{L}) we find:
\begin{eqnarray}
h_{12} & = & \left( \frac{a^3 a^{'*}}{M_1^2} + \frac{a |a^{'}|^2 a^{'*}}{M_1^2}
\right) + \frac{a a^{'} b^2}{M_1M_2} \nonumber \\
h_{23} & = & \left( \frac{b b^{'} a^{'2}}{M_1 M_2} \right)
 + \left( \frac{b^3 b^{'}}{M_2^2} + \frac{b b^{'3}}{M_2^2} \right)
\nonumber \\
h_{31} & = & \left( \frac{a a^{'*} b b^{'}}{M_1 M_2}
\right) \nonumber \\
\end{eqnarray}
from which it follows that
\begin{equation}
\xi_L =  - \frac{a^6 b^6}{M_1^3 M_2^3} sin 2\delta [ Y^2 (2 + Y^2)]
\label{lowenergy}
\end{equation}
Here we have taken $b=b^{'}$ because
the mixing for the atmospheric neutrinos
is almost maximal.

Neutrinoless double beta decay $(\beta\beta)_{0\nu}$
is predicted at a rate corresponding to $\hat{L}_{ee} \simeq 3 \times 10^{-3}eV$.

The comparison between Eq.(\ref{highenergy})
and Eq.(\ref{lowenergy})
now gives a unique relation between the signs of $\xi_L$ and $\xi_H$.

This fulfils in such a class of models the idea of \cite{Sakharov}
with only the small change that baryon
number violation is replaced by lepton number violation.

\section*{Acknowledgement}

I wish to thank Ruth Daly, Tom Kephart, 
the late Behram Kursunoglu, 
Steve Mintz, Arnold Perlmutter and Ina Sarcevic
for organizing the very memorable Fest. 
This work was supported in part
by the Department of Energy
under Grant Number
DE-FG02-97ER-410236.

\end{document}